# Lack of water and endurance running could have caused the exponential growth in human brain
## "Point of no return" model

**Konrad R. Fialkowski**
University of Warsaw, Poland; An den Langen Luessen 9/1/3; 1190 Vienna, Austria;
e-mail: fialkows@aol.com

Growth in brain volume is one of the most spectacular changes in the hominid lineage. The anthropological community agrees on that point. No consensus, however, has been reached on selection pressures contributing to that growth. In that respect Martin (1984) can be invoked. In his review of size relationships among primates he stated that despite the relationship between brain size, body size and feeding behavior no single interpretation could be provided that revealed the causality of such relationship.

This paper deals with one specific aspect of hominid brain growth: the fact that for most of the hominid period, growth in brain volume was exponential in character. To the author's knowledge, no attempt has been made to identify a selection mechanism that can facilitate just the exponential features of that growth (as distinct from any of its other characteristics). It is broadly accepted that the dynamics of this growth were peculiar. Growth was very fast, or even rapid in the evolutionary scale of time. The most profound evidence of that opinion was expressed by Haldane (quoted after Mayr 1970: 384): "J. B. S. Haldane liked to emphasize that this dramatic increase in brain size was the most rapid evolutionary change known to him".

## EXPONENTIAL GROWTH OF HOMINID BRAIN VOLUME

In mathematical terms the exponential function is the very fast growing function. Funds deposited in a bank for several years grow exponentially on account of the interest being added to it each year, thus increasing the value of the interest in consecutive years. The number of stones in a landslide grows exponentially, when statistically each stone initiates movement in more than one other stone. The number of neutrons in a nuclear chain reaction also grows exponentially. For each of these examples, it is relatively easy to discern a mechanism behind the exponential growth.

That notwithstanding, no obvious reason present itself for this type of brain volume growth in the hominid lineage. First and foremost, it has to be demonstrated that this growth was indeed exponential, or more precisely, that the relationship between brain size in fossil hominids and time is best approximated by the exponential function. Most researchers directly or indirectly confirm this. For example, Stringer (1984, quoted after Foley 1987: 149) presented this relationship as a straight line on the diagram with a linear scale of time (x) and a logarithmic scale of volume (y). It implies the exponential character of the function. Using data from Tobias (1987), Bickerton (1990: 133- 136) offered an analysis of the brain growth curve that indicated its exponential character. Grüsser (1990: 356 – 359) explicitly presented equations that described the exponential character of the brain growth function. Moreover he also presented (ibid.) the function describing the velocity of brain growth (first derivative of the brain growth function). According to his description the function of brain growth was exponential from the onset of growth until approximately 200,000 years BP, when the inflection (turning) point occurred. He also states (ibid. 353) that empirical data do not support occurrence of the "punctuated equilibrium", as far as brain growth is concerned.

Mathematically, the exponential growth of brain volume means, that in **each** generation, average brain volume is x times larger than average brain volume in the preceding generation, where x is a real number larger than 1. Certainly, this mathematical description is a simplification, since in reality, the value of x usually differs from generation to generation. The constant value of x solely approximates the growth trend over many generations.

In the terms of natural selection, this would mean that in each generation the same percentage of individuals with the smallest brain is excluded from the reproduction process. This percentage is the same **independent** of the average brain volume already achieved. Thus, the population in the consecutive generation faces a situation known as the Red Queen syndrome; it "runs" towards the larger brain, as fast as it "can" (exponentially) and selection-wise it remains exactly where it was, since despite the progress in adaptation, the same percentage is negatively selected. This means that the progressive adaptation (increase in brain volume) does not relax selection pressure.

This is an extremely strange arrangement for natural selection. For non-biological selection, similar cases of non-relaxed selection could be identified (not necessarily, however, leading to exponential growth, for which other conditions also have to be met). A case in the point is the Olympics Games and the results in the most objectively measurable competitions such as track-and-field events. The ever better records set in those disciplines are the result of selection governed by artificial rules. According to those rules, only three places are selected positively. Competition for those places drives the contestants to ever greater heights and in most cases the gold medalists from the early days of the Olympic Games would have no chance whatsoever of winning a medal today.

A change, which is exponential in character, is the result of positive feedback. It means that progress in adaptation results in increased selection pressure being caused by that progress. In other words, as adaptation progresses more and selection pressure increases apace, the net result is non-relaxed selection pressure for continuously advancing adaptation, i.e. precisely the Red Queen syndrome.

For example, a causality chain: larger brain $\Rightarrow$ more efficient hunting $\Rightarrow$ resultant decrease in prey might satisfy requirements for positive feedback. This chain, however, contradicts known facts. According to the requirements for the positive feedback applied to this case, the availability of prey should decrease **monotonically** - at least between 1.6 MYBP and 0.2 MYBP. To meet those requirements, a steadily increasing overkill of **all** potential hominid prey would have had to continue throughout the period of approximately 2 million years; this did not occur. For example Bortz (1985: 148) noted: "A comparison of the kills made by a group of contemporary Bushmen was remarkably similar to the bony remnants from Olduvai (Spaeth & Davis, 1976)".

Although the emergence of *Homo erectus* apparently coincides with a major shift in the predator/prey system of large mammals (Walker, 1984), the shift is invoked as an argument for the evolutionary and ecological instability at that time facilitating the hominids' entry into "the guild of large carnivores" (Foley, 1987: 260). The arrangement needed to justify the existence of the feedback discussed would have to be a shift that resulted from the hunting of *Homo erectus* and one that was continuously enhanced throughout the whole period of exponential growth. Moreover, not the shift alone, but exclusively the monotonically decreasing prey availability caused by such a shift might have justified the existence of the above-mentioned feedback.

Nevertheless, for exponential growth feedback in the process may always be identified.

One essence of selection is competition forced by limitations. In a search for feedback (applied later to the model presented here), an attempt was first made to identify limiting factors in hominid lineage that an increase in brain volume could not have overcome before 0.2 MYBP. Furthermore, such factors should:

(i) Remain unchanged over the time span of exponential growth (i.e. approximately 2.0 - 0.2 MYBP);

(ii) Be independent of brain growth; and

(iii) Be superimposed on the progressive adaptation of the increase in brain volume and thus cause the positively selected segment of the population to have been roughly the same in percentage terms throughout the whole process of exponential brain growth (analogous to the rule limiting the award of Olympic medals to the first three past the post).

## WATER DEPENDENCE MODEL

A factor that best fits the requirements listed above is **water dependence** in hominids.

Foley (1987: 106) describes this dependence as follows:
"Modern humans can withstand only limited water loss (up to 10 per cent of body weight), and are unable to ingest large quantities of water (1 liter per 10 minutes, compared with 100 liters per 10 minutes for camels). The principal consequence of sweating is the need for hominids to keep close to water. (...) ...hominids, if they were as water-dependent as modern humans and most primate species, will be **limited to areas with permanent surface water. With savannah environments these may be highly localized. (...) The non-focal foraging patterns within a home range of many species would not be appropriate**." (emphasis added).

In drawing up a model based on water dependence, a single source of surface water surrounded by dry savannah was assumed. It was also assumed that the hominid population had a (permanent or movable) home base in close proximity to that source.

Chasing prey (persistence hunting: Krantz 1968; Watanabe 1971; Carrier 1984; Bortz 1985; also quoted after Carrier: Schapera (1930); Bennet & Zingg 1935; Lowie 1924; Foster 1830; Solas 1924; McCarthy 1957) was a prevailing method of hunting in early hominids and despite the adoption of projectile weapons and other technologies by modern humans it is still in use today. The motives behind maintaining persistence hunting in contemporary human populations were presented by Bortz (1985:147):

"When prey density is low, individual hunting is wisest (Lamprecht, 1978). To obtain highest return per amount of time and energy expended in searching for a mobile resource, the best strategy would seem to be cover as much area as possible per person" (Hayden, 1981).(...) chase myopathy renders any animal incapable of further retreat or defense so that individual hunting may have been very effective, indeed it could have been the predominant behavior. Hayden wrote, "Groups will hunt as individuals when they can and communally when they have to" (Hayden, 1981)".

It should be noted, however, that when hominids joined "the guild of large carnivores" those newcomers apparently had no genetically established group hunting strategies. Without speech, determining and exchanging details of plans to be executed **on the open savannah** may have been beyond the hunters' capabilities (including their mental capacity). The limited extent of those capabilities is evident in the lack of any substantial progress being achieved in tool production until late *Homo erectus* (Wynn, 1988).

Chasing prey results in heat stress and increases consumption of the body's water resources. With the single water source in a savannah environment assumed for the model, hominid hunters in pursuit of their prey would not have been able to reach (by definition of the model) other water sources on the savannah. With the fluid resources of their bodies exhausted through running, they would have collapsed and died before reaching the water source. They might possibly have reached other water sources by walking, given that fluid loss through sweating is less than when running. For the model, however, this is of secondary importance.

When running, each individual can cover a maximum distance from the water source to which return by walking is possible, without the body water resources of that individual being completely exhausted. All possible routes fulfilling this condition are within a circle surrounding the water source. Its radius is equal to that maximum distance. For an individual, the circle so determined encompasses all the 'points of no return' for that individual. Akin to a plane with only one fuelling point, a return to the starting point from beyond the 'point of no return' circle is impossible. Pursuing prey beyond the circle means death for the hunter.

Hunting effectiveness is assumed to be positively correlated with brain volume. This means that in a time span that commences with the pursuit of the prey and ends at the latest when the maximum distance from the water source has been reached, the probability of success is higher for individuals with larger brain volume in comparison to hunters of smaller brain volume.

It should be noted that **independent of the total availability of prey, the prey resources available for hunters are always limited and restricted to the prey within the 'point of no return' circle.**

Once the hunt starts, it must finish before the maximum distance is covered. Those hunters who extend the hunt beyond that range at the expense of the body water resources they need for their return are usually lost (unless the hunt is ultimately successful beyond the 'point of no return' and the hunter can draw on the prey's blood to supplement body water resources).

From the selection conditions (i.e. from a positive correlation between brain volume and hunting success) it follows that those hunters who went missing were statistically those with the smallest brain, since those with larger brains had a greater chance of success before reaching the 'point of no return'.

Given the limited prey resources within the 'point of no return' circle' the hunters compete indirectly among themselves. This indirect competition constitutes a selection pressure for brain volume growth that increases hunting success. This selection pressure over a number of generations and longer time spans is independent of many other selection factors. For example, an increase in total prey availability that also results in an increase in prey availability within the circle relaxes competition only temporarily. Increased food supply would be followed by an increase in numbers of the hominid population and competition would eventually revert to its previous intensity. A decrease in the supply of prey would immediately increase competition, strengthen selection, decrease population size and finally (less prey, fewer hunters) lead to competition among the hunters similar to the initial state.

Because selection pressure results mainly from competition among hunters, the strength of this selection pressure does not depend on the brain volume value already achieved in the population. Independent of its current value, the survival chances of those individuals with the smallest brain will always be lower than those hunters with larger brains. In a sense, selection pressure in the model is invariant to brain volume.

It is also invariant to different running capabilities or body size in different hominids. With the earliest hominids' inferior running capabilities the radius of the 'point of no return' circle is smaller. It may decrease the number of individuals in those early hominid populations in comparison to the ones that followed later; it does not, however, substantially influence the force of the selection pressure, which is mostly the outcome of indirect competition among the hunters.

The model was developed for persistence hunting where success is dependent on individual, uninterrupted pursuit of the prey. For the validity of the model, it is crucial that the hunters lack two capabilities:
 (i) They cannot carry water with them; and
 (ii) Their mental resources are insufficient to estimate the point of no return (humans provide an example of the lack of genetically based estimations in that respect. Such estimations are performed on the basis of advanced mental capabilities in humans).

The model is also valid for more than one single water source, unless the distribution of the sources is too dense. For two water sources located closer to each other than twice the hunter's maximum range, the contour line linking the 'points of no return' is more complicated than the circle. Under these circumstances, the probability of hunting success depends on the location of the hunter within that contour. With the increase in the number of water sources, selection pressure decreases as the probability of overshooting the 'point of no return' lessens.

The maximum range varies among individuals. For the population as a whole, the 'point of no return' circles around a single water source drawn together for all the individuals in that population would constitute a ring. Inside that ring there is a white circular zone, whence any individual can return to the water source. The ring, surrounding the white zone, is a grey zone containing the 'point of no return' circles for all the individuals in that population. Outside that ring the black zone begins, from which no individual can return. For the correctness of the model discussed, the grey ring should be narrow. The model can thus be falsified. The model strongly implies one particular feature of hominid adaptation to endurance of the limited availability of water. As long as brain volume continued to grow exponentially, the hominid genotype had to be fixed for genes determining endurance in relation to water. The variability of genes determining water-specific endurance in the population had to be nil or very close to nil. Despite 200,000 years of relaxation of selection pressure on water endurance, genetic drift could not have drastically changed the variability of human genome in that respect. This variability **must not** have been normally distributed. A substantial part or majority of the human population should have preserved to this day this genetically determined endurance at its highest level, as achieved by hominids, whereas only some individuals may have partially lost some of the genes needed for this highest level of endurance.

## DISCUSSION

This model is supplementary to the proposal that heat stress was the selection factor in hominid evolution (Fialkowski, 1978, 1986). Although it was developed for that proposal, its applicability is determined solely by the assumptions of the model. This means that when the assumptions for the model are met, a positive correlation between brain volume and effectiveness of persistence hunting is sufficient for its applicability.

According to the proposal (Fialkowski, 1978, 1986) hominid brain emerged as a result of preadaptation. In preadaptation, a structure emerging as a result of a selection pressure is, by chance, appropriate for a new function that differs from the one which originated the selection pressure. Apart from adaptation, it is the second possibility offered by the Darwinian theory. According to Mayr (1970: 423) a structure is preadapted, if it can assume a new function without interfering with the original function. A reliability hypothesis (Fialkowski 1978) first regarded preadaptation as a mechanism for the origin of a large and highly interconnected human brain. The hypothesis claims that:

(i) heat generated in hominid bodies during persistent hunting/running (Krantz 1968; Bortz 1985) was transported from the muscles via the blood stream to the brain, damaging neurons at random, impairing brain functions and decreasing hunting success. Effective blood cooling systems as in other mammals (Baker 1972, 1979; Baker & Chapman 1977) were not developed.

(ii) In terms of hunting success, the number of malfunctioning neurons in the brain tissue was irrelevant as long as the brain continued to function properly. Any variations in the brain structure, which increased the capability of the brain to maintain its function as a whole, despite some malfunctioning neurons, were strongly positively selected.

**(iii)** A reliability principle (von Neumann, 1963) states that in order to increase the reliability of information structure composed of malfunctioning elements, both the number of elements and the number of connections between the elements must be increased. **The reliability hypothesis claims that this principle found by von Neumann constituted a pattern for brain adaptation in hominids.**

As a result of this adaptation, the adapted brain should (Fialkowski 1990b):
(i) have an increased number of neurons;
(ii) the neurons should be more interconnected; and
(iii) the brain should be more resistant to heat stress.

Both features, (i) and (ii), deduced from von Neuman's theory are specific to reliability adaptation and can be found in the human brain: interconnectivity is greater than in the brain of great apes, and phylogenic growth of the brain volume is a manifestation of certain increase (1.25 times; Holloway 1966) in the number of neurons, which are less densely packed (Shariff 1953). The third feature, (iii), is also specific to the human brain. The human brain is clearly more resistant to heat stress than that of animals. As Brinnel *et al*. (1987: 209) put it (emphasis added):

"…in a view of the high levels of body temperature which have been recorded in runners (about 42° C) or in heat stroke patients (46.55° C), either there is a very appreciable extent of selective brain cooling or **the brain is much less temperature sensitive than indicated by animal experiments**."

More detailed discussion of the subject could be found in Fialkowski (2013) and justification for the whole approach in Fialkowski and Bielicki (2008).

In the context of the discussed model, the 'Machiavellian intelligence' (1988) approach should be compared. Machiavellian Intelligence also implies positive feedback. The model presented here does not contest the Machiavellian intelligence approach. It seems that Machiavellian intelligence as a selection mechanism for hominid lineage is applicable as a primary selection mechanism after 0.2 MYBP rather than earlier. Machiavellian intelligence assumes selection mostly through social selection pressures. It implies selection through differential fertility or a group selection rather than through differential mortality. Both of those social-selection pressures are rather weak: being exercised within, rather than outside the group. The exponential growth is a result of extremely powerful selection (e. g. the opinion of Haldane quoted earlier). It implies relatively drastic selection pressures. Implementation of such strong selection pressures through differential fertility means sex monopoly. Moreover, this sex monopoly had to be both exercised in a co-operative group and positively correlated with brain volume (and not with an individual's strength, for example). That is hardly possible.
Differential mortality pressure exercised inside the hominid group is also difficult to accept, taking into consideration the apparent food-sharing feature displayed by hominids groups.

Generally, drastic selection could for the most part be exercised via differential mortality. As for hominids, selection of that kind could have been exercised outside the group rather than within. It is indicative of selection during hunting and/or gathering activities. Heat stress as a primary selection factor (Fialkowski, 1978, 1986, 2013) fits this pattern well. The main source of heat, however, is a by-product of physical activity rather than sole exposition to sun radiation in a hot environment. As Bortz (1985: 148) stated it is: "...the heat generated by exercise which is the discriminating burden". Thus, hunting, especially persistence hunting, fulfils the requirements for the behavior under drastic selection conditions that is required for the rapidly progressing adaptation discussed.

As I have attempted to justify in this paper, by its very character adaptation exponential is more the outcome of internal competition among individuals in the population than something driven by outside factors. On the other hand, however, extremely strong direct internal competition cannot be exercised within a co-operative group. This contradiction can be avoided solely by indirect competition, such as that outlined in the model. It may remain strong and at the same time does not contradict co-operation within the group.

Exponential growth in brain volume came to a conclusion in approximately 0.2 MYBP. Apparently, it was the emergence of speech that brought an end to exponential growth. With the advent of symbolic communication, other, more sophisticated patterns of co-operative hunting could be introduced. Repeating the quotation of Hayden (1981, after Bortz 1985): "Groups will hunt as individuals when they can and communally when they have to". Given the faculty of speech, the group (contrary to earlier hominids) could have hunted "communally when they [had] to" i.e. during difficult times when prey was scarce. It was precisely these difficult times that constituted the period of strongest selection (Foley, 1987). Thus, the selection pressure that had become too strong while hunting individually was relaxed via a more sophisticated hunting mode available to hunters after the emergence of speech. As a result, brain growth ceased to expand exponentially and the inflection point occurred on the exponential curve.

An independent dating of the emergence of fully developed speech between 125000 and 250000 BP was given by Liberman (1991: 109, 250 respectively). It coincides with the inflection point. It confirms a prior prediction (Fialkowski 1990: 188) derived from the heat stress hypothesis. In line with this prediction, it was proposed (Fialkowski 1994) that the emergence of speech had been preceded by language-oriented brain structures and that both phylogenetically (ibid.) and ontogenetically (Fialkowski & Szymanski 2000) speech could not have emerged prior to consciousness. If those proposals were correct, they imply that speech, not consciousness alone, was the milestone in hominid evolution. Its emergence concluded the exponential growth of brain volume in hominids and generated new social selection pressures in addition to enhancing those that already existed (Machiavellian Intelligence, 1988; Dunbar 1993) that were followed by new adaptations and ultimately culture.

KEY WORDS: heat stress, water dependence, reliability of information structure, von Neumann's principle, brain volume, human origin.

ACKNOLEDGEMENT: The author thanks Professor B. K. Szymanski for their comments concerning the text.